\documentclass[multphys,vecphys]{svmult}

\usepackage{makeidx}         
\usepackage{graphicx}        
\usepackage{multicol}             
\usepackage[bottom]{footmisc}   
\makeindex

\begin{document}

\title*{Spectropolarimetry of the Deep Impact target comet 9P/Tempel 1 with HiVIS}
\titlerunning{Spectropolarimetry of Deep Impact}
\author{D.M. Harrington \inst{1} \and K. Meech \inst{2} \and L. Kolokolova \inst{2} \and J.R. Kuhn \inst{1}  \and K. Whitman \inst{1} }

\institute{University of Hawaii, Honolulu, HI  96822
\texttt{dmh@ifa.hawaii.edu}
\and University of Maryland, College Park MD 20742}

\maketitle


\setcounter{footnote}{0}

\begin{abstract}

Spectropolarimetry of the Deep Impact target, comet 9P/ Tempel 1, was performed during the impact event on July 4th, 2005 with the HiVIS Spectropolarimeter and the AEOS 3.67m telescope on Haleakala, Maui. We observed atypical polarization spectra that changed significantly in the few hours after the impact.  The polarization is sensitive to the geometry, size and composition of the scattering particles.  Our first measurement, beginning 8 minutes after impact and centered  at 6:30UT, showed a polarization of 4\% at 650 nm falling to 3\% at 950 nm.  The next observation, centered an hour later, showed a polarization of  7\% at 650 nm falling to 2\% at 950nm.  This corresponds to a spectropolarimetric gradient, or slope, of -0.9\% per 1000\AA\ 40 minutes after impact,  decreasing to a slope of -2.3\% per 1000\AA\ 75 minutes after impact.   Both are atypical blue polarization slopes.  The polarization values of 4\% and 7\% at  650nm are typical for comets at this scattering angle,  whereas the low polarization of 2\% and 3\% at 950nm is not.  This, combined with the IR spectroscopy performed by a number of observers during the event, suggests an increase in size, number, and crystallinity of the individual silicate particles (monomers) that are a constituant of  the dust particles (aggregates) in the ejecta.  

\end{abstract}

\section{Introduction}
\label{sec:1}

	The angular and spectral dependence of coma polarization are signatures of the size, composition and structure of coma grains \cite{hanner02, hadamcik03c, kolokolova04}.  The degree of polarization typically increases with wavelength, called a red polarimetric color for the dust coma, as well as phase angle \cite{kolokolova04, kolokolova97, lev03}.  A few comets have shown a blue polarimetric color such as comet 21P/Giacobini-Zinner \cite{kiselev00}.  The polarization and polarimetric color are influenced by the size, composition, and geometry of the individual scattering monomers that make up an aggregate dust particle.  On 2005 July 4 at 5:52:02 UT comet 9P/Tempel 1 was impacted by a 364 kg spacecraft as part of the NASA Deep Impact mission \cite{ahearn05}.  The event presented a unique opportunity to compare spectropolarimetric measurements with other measurements performed simultaneously as part of the Deep Impact mission to further constrain the dust properties.

	For comets at phase angle of 40.9$^\circ$, average coma polarization in a red filter is 4\% to 10\% and perpendicular to the scattering plane plane, usually increasing towards 1$\mu$m \cite{lev03}.  Some exceptions are known however \cite{kiselev00, lev99, lev03}.  Some dependence on comet activity was also found and explained as a consequence of the influence of scattering particles of differing size and composition \cite{hadamcik03a, hadamcik03b, hadamcik03c}.  A change in polarization due to a change of composition and size is also seen in scattering models \cite{kolokolova97, kolokolova04, kimura06, las06}.
 
	The polarization and color of cometary dust scattering has been studied theoretically and experimentally in great depth over a great range of variables such as size, composition, scattering geometry, particle roughness, and aggregation process \cite{gus99, kolokolova04, kimura06, las06}.  Models that fit the typical angle and red wavelength dependence of the polarization best use aggregates of more than 1000 monomers having small radii ($\sim$100nm).  Model aggregates that fit the typical polarization measurements have a high index of refraction (m=n + ik, n=1.8 to 2.0, k$\sim$0.4 to 0.6).  This corresponds to volume fractions of one-third silicates, two-thirds carbonaceous materials, and a small amount of iron-bearing sulfides \cite{kimura06}.  This material is quite dark (high imaginary part of the complex index of refraction) and similar to the composition of 1P/Halley's dust  \cite{mann04}.  The carbonaceous material in the model is roughly two thirds amorphous carbon and one third organic-refractory material.  The type of aggregation (cluster-cluster vs. particle-cluster) did not play a major role in the polarization models, but it does influence the thermal IR spectra through the porosity \cite{kimura04, kolokolova04}.  A decrease in polarization with wavelength, called blue polarization gradients or slopes, were not extensively modeled because of their rarity.  However, some models demonstrated the tendency of polarimetric color to decrease, {\it i.e.} get more blue, with increasing size, albedo ({\it e.g.} icy), and transparency of the material (crystallinity), whereas more absorbing materials typically show an increase in polarzation with wavelength, called a red polarization slope \cite{kimura06}.  Multiple scattering in optically thick dust clouds depolarizes the scattered light.  With all this information contained in the polarization of scattered light, we expected the polarimetric data to shed light on the change in dust properties resulting from the Deep Impact encounter.

\section{The Deep Impact Spectropolarimetry}

The AEOS telescope is a 3.67m, altitude-azimuth telescope.  The HiVIS spectrograph is a cross-dispersed echelle spectrograph using the f/200 coud\`e optical path \cite{thornton03}.  Since non-normal incidence reflections change the polarization state of the incident light, a careful calibration of the telescope has been performed \cite{harrington06}.  We used a 1.5$''$$\times$7$''$ slit (970$\times$4530 km at a geocentric distance of $\Delta$=0.89 AU) and the ''red" setting (nominally 637.5-968.0 nm) for all comet observations.  The spectropolarimetry module for the AEOS spectrograph consists of a rotating achromatic half-wave plate and a calcite Savart Plate. The Savart plate separates an incoming beam into two parallel, orthogonally polarized beams separated in the spatial direction at the focal plane.  

A common definition in planetary science is that the degree of polarization is the difference between the intensity polarized parallel and perpendicular to the scattering plane \cite{kolokolova04}.  Since the HiVIS image rotator was not used in order to simplify the polarization calibration, the projection of the slit's position angle onto the sky was not constant, and knowledge of the scattering plane is difficult to extract.  In this paper, we will be using an alternative definition that does not require knowledge of the scattering plane, simply calculating the polarization in the instrumental reference frame (see \cite{harrington07} for details).

\begin{figure}
\centering
\includegraphics[height=14cm, angle=90]{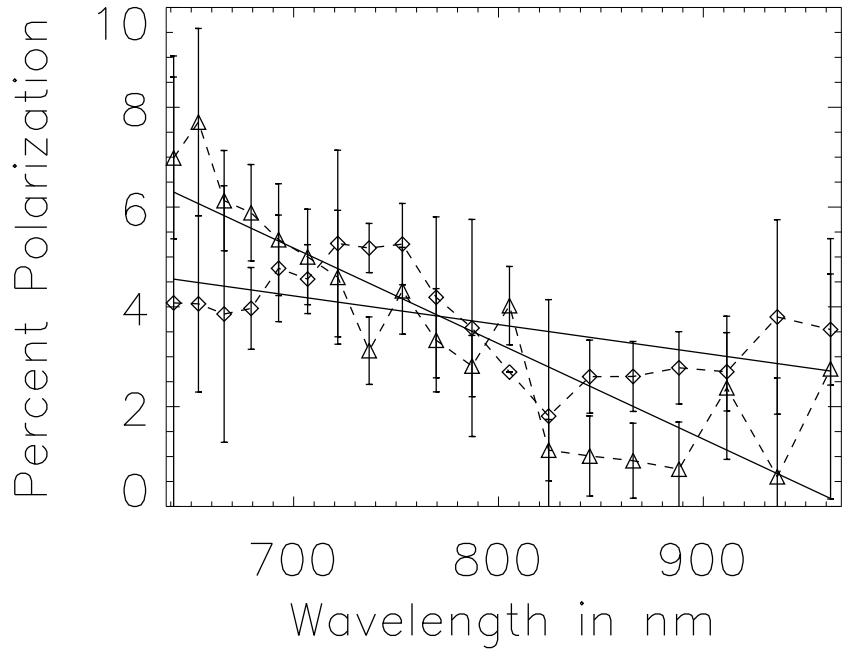}
\caption{Polarization for comet 9P/Tempel 1 with 3-$\sigma$ error bars.  The data was taken on impact night from 6-7 UT (diamonds) and 7-8 UT (triangles) at a phase angle of 40.9$^\circ$.  A linear fit is plotted to guide the eye.  A single point at 805nm (order 11) in the 6-7 UT data set has been replaced by the average of neighbor points, and has no error bars.  The polarization spectra from 6-7 UT curve had a shallow negative slope of -0.9$\pm$0.2\%/ 10$^3$ \AA. The 7-8UT curve had a slope of -2.3$\pm$0.3\% / 10$^3$ \AA.}
\label{fig:1}       
\end{figure}

Comet 9P/Tempel 1 was bright enough to obtain useful spectropolarimetry only on the night of impact.  We aquired two complete data sets, from 6-7 and 7-8 UT.  The data were reduced using the AEOS-specific reduction pipelines developed for this instrument \cite{harrington06}.  Calibration of comet spectropolarimetry is performed by interpolating a sky-map of unpolarized standard star observations to the pointing of the comet to create a telescope-induced polarization calibration.  The corrections were typically 2-3\% with mild wavelength dependence ($\leq1$\%) and they did not alter the slope-change seen in the comet 9P/Tempel 1 \cite{harrington07}.  In order to present high signal to noise measurements of the polarization, the spectra were averaged 1000:1, giving 19 independent polarization measurements.  The spectropolarimetry is plotted in Fig. \ref{fig:1}.  The 6-7 UT data set, started 8 minutes after impact, shows a slightly anomalous blue-sloped degree of polarization of 4\% falling to 3\% from 650 to 950nm (-0.9$\pm$0.2\%/10$^3$\AA).  In contrast, the 7-8UT data set, started 75 minutes after impact, shows a more pronounced blue slope from 7\% at 650nm to 2\% at 950nm (-2.3$\pm$0.3\%/10$^3$ \AA). This is an indication of the change in particle scattering properties.  The leading edge of the ejecta plume was moving out from the nuclues at $\sim$200m s$^{-1}$ \cite{meech05}.  A body moving this speed could move across the slit in 40 minutes, setting the timescale for significant change in scattering by ejected gas and dust, consistent with other observations \cite{meech05, jehin06, sugita05, schleicher06}.     

For comparison, Comet C/2004 Q2 Machholz was observed on November 27, 2004 at a phase angle of 30$^\circ$.  Two complete data sets (8 images at 1200s, roughly double the flux) were reduced and calibrated in the same way as the comet 9P/Tempel 1 data.  The measured degree of polarization was in the usual range and showed a typical red polarization slope which did not change very signifigantly between the two image sets.  This result gives us confidence in our instrument calibration and reduction techniques.

\section{Discussion and Conclusions}

Our observation of 4\% and 7\% polarization at 650nm is typical for comets at these wavelengths and phase angles, 4\% being somewhat low.  However, the 1\% and 3\% polarization at 950nm is not at all typical. The few computer or laboratory simulations that have modeled blue polarization slopes show it can result from larger particles or a predominance of transparent particles (larger crystalline silicates or ices) \cite{hadamcik02, kolokolova04, kimura04, kimura06, las06, kiselev00}.  At this point, we must discuss other observations of silicates, ices, and particle sizes to fully interpret the polarization.

The infrared observations of comet 9P/Tempel 1 right after the impact showed a strong and complex silicate feature developing by 1 hour after impact, fading 1.8 hours after impact  \cite{harker06, sugita05, lisse06}.  The size of the silicate particles also evolved strongly.  The dust particles were reported to have size distribution peaks (of the aggregates, not monomers) increasing after impact \cite{harker06, lisse06, lisse06s}.  These models showed that pre impact there was an absence of sub-$\mu$m silicates, and the spectrum was mineralogically dominated by larger (0.9$\mu$m) amorphous olivine with no carbon or crystalline silicates.  They reported an increase in the number of sub-$\mu$m silicates, an emission from relatively transparent Mg-rich crystalline olivine, as well as a doubling of the silicate to carbon ratio and a 4-fold increase in the crystalline to amorphous silicate ratio between the first and second hour after impact.  A relative lack of organics was also seen with the amorphous carbon being roughly 20\% of the silicates (30\% to 50\% is typical) \cite{lisse06}. 

There was evidence for ice, but in a smaller amount than necessary to produce a blue polarization slope.  There was a change in color of the dust in the near-IR post impact interpreted as icy grains, or icy grain mantles being liberated in the impact event \cite{fernandez06}.  There was direct observation of icy grains in the ejecta plume from the main spacecraft in the form of the 3 $\mu$m ice absorbtion from immediately after impact through lookback 46 min later \cite{sunshine06}.  Evidence for icy grains in the inner 600km of the coma was seen with the sublimation maximized 1.5 hours after impact \cite{schulz06}.  Particle disintegration was suggested from the different spatial evolution of CN, [OI], and dust continuum flux in spectrophotometric measurements \cite{hodapp06}.  The Spitzer spectra also required ice covering 3\% of the dust surface area in their 10" FOV \cite{lisse06}.  Thus, ice is also a possibility as a contribution to the transparency of the particles.  However, the amount of ice necessary to dominate the scattering is much more than is suggested by the Spitzer and spacecraft data.

Multiple scattering depolarizes light in cometary dust and may be responsible for the low polarization just after impact.  An optically thick plume was seen, thinning to $\tau\sim0.4$ 20-25 minutes after impact \cite{schleicher06}.  The look-back images from the fly-by spacecraft also showed an optically thick ejecta plume after impact \cite{ahearn05}.  Since the optical depth was a strong function of time, the influence of multiple scattering is assumed to change strongly as well.

We can explain our observations by a depolarization due to multiple scattering in the first hour and subsequent domination of larger and more transparent monomers (silicates), and possibly ices, in the ejected dust aggregates of comet 9P/Tempel 1.  This is consistent with the infrared data on Deep Impact, which indicate a high amount of silicates in the DI dust whereas in situ data for comets 1P/Halley and 81P/Wild 2, typical red polarization comets, show that their dust contained two-thirds of carbonaceous materials and organics \cite{lisse06, harker06, sugita05, kimura06, kis04}.  We can speculate that subsurface materials in comet 9P/Tempel 1 had more volatile organics and ices that quickly evaporated or decayed leaving the disrupted and fragmenting dust, rich in silicates, to produce the observed blue polarization slope \cite{fernandez06, mumma05, ahearn05}.  Harker et al. have suggested that smaller particles in the ejecta moved faster, size sorting the cloud, leaving larger and more crystalline silicates behind \cite{harker06}. This could also contribute to the anomalous blue polarization slope we detected for the inner coma.

\printindex
\end{document}